\documentclass[aps,prb,superscriptaddress,twocolumn,floatfix,showpacs,amsmath]{revtex4-1}
\usepackage{graphicx}
\graphicspath{ {./images3/} }
\usepackage{floatrow}
\usepackage{latexsym}
\usepackage[makeroom]{cancel}
\usepackage{dcolumn}
\usepackage{epstopdf}
\usepackage{float}
\floatstyle{plaintop}
\restylefloat{table}
\usepackage{color}
\usepackage{amsmath,latexsym,psfrag}
\usepackage{array}
\usepackage{dcolumn}% Align table columns on decimal point
\usepackage{bm}% bold math
\usepackage{epstopdf}
\usepackage{multirow}
\usepackage{soul}  % \st    for strike-out
\usepackage[
colorlinks=true,
urlcolor=blue,
linkcolor=blue,
citecolor=blue]{hyperref}

\begin{document}

\begin{titlepage}

\title {An insight into the properties of ATiO$_3$ (A=Ti,Sr) materials for photovoltaic applications}

\author{Lynet Allan}
\email{Corresponding author: allanlynet3@students.uonbi.ac.ke}
\affiliation{ Department of Physics, School of Physical Sciences, University of Nairobi, P.O.Box 30197-00100 Nairobi Kenya.}

%\author{Winfred M. Mulwa}
%\email{winfred.mulwa@egerton.ac.ke}
%\affiliation{ Department of Physics, Egerton University, P.O Box 536-20115 Egerton Kenya} 

\author{Julius M. Mwabora}
\affiliation{ Department of Physics, School of Physical Sciences, University of Nairobi, P.O.Box 30197-00100 Nairobi Kenya.}

\author{Musembi J. Robinson}
\affiliation{ Department of Physics, School of Physical Sciences, University of Nairobi, P.O.Box 30197-00100 Nairobi Kenya.}

%\author{ Barnard Aduda O.}
%\affiliation{ Department of Physics, School of Physical Sciences, University of Nairobi, P.O.Box 30197-00100 Nairobi Kenya.}

\date{\today}

 \begin{abstract}
One of the most active research areas in the world is the search for effective materials for use in the fields of optoelectronics and photovoltaics. The potential of materials like ATiO${_3}$  (A=Ti,Sr) is yet largely untapped. Ab initio studies based on density functional theory (DFT) have been used to comprehensively explore the structural, electronic, elastic, and optical properties of Ti$_2$O$_3$ and SrTiO$_3$. In this study, the ground state properties were computed with spin-orbit coupling (SOC), without spin-orbit coupling, and with the inclusion of Hubbard U parameter. Ti$_2$O$_3$  has been found to have electronic bandgaps of 0.059 eV without SOC, 0.131 eV with SOC, and 1.665 eV with Hubbard U. For SrTiO$_3$, electronic bandgaps of 1.612 eV, 1.761 eV, and 2.769 eV have been obtained, respectively, without SOC, with SOC, and with Hubbard U. Ti-4d orbitals have been observed to dominate near the top of the valence band in each and every instance. SOC did not significantly affect the bandgaps and calculated lattice parameters for ATiO$_3$  (A=Ti,Sr). ATiO$_3$  (A=Ti,Sr) is mechanically stable at absolute zero pressure, according to the mechanical stability test. The optical band gap has been seen to increase when Hubbard U is taken into account. In general, the Hubbard U parameter enhances bandgap and optical property predictions. Ti$_2$O$_3$ and SrTiO$_3$ are good UV-Vis absorbers and appropriate for photovoltaic applications owing to the optical absorption coefficient curves being found to cover the ultraviolet to visible (UV-Vis) region.
 

\end{abstract}
\maketitle
\end{titlepage}

\section{Introduction} \label{Sec :I}
The search for sustainable energy resources has become a key field of study due to rising global energy demands and the negative environmental pollution generated by the burning of fossil fuels \cite{eddiouane2018first,samat2016hubbard,umebayashi2002analysis}. Titanium based oxides have been under investigation for the past few decades owing to their wide range of applications in environmental remediation and solar energy conversion,\cite{samat2019structural,umebayashi2002analysis,giustino2014materials}. Titanium is one of the earth-abundant elements, and its oxides including titanium dioxide (TiO$_2$ ) and ATiO$_3$ (A= Ba,Ca,Fe) have gained significant attention for applications in technologies of electronics, energy conversion, catalysis, sensing, and so on\cite{zhang2010promising,anderson1963simplified,inturi2014visible,li2018titanium}. Despite its large band gap, TiO$_2$ has been researched for its potential modification for solar harvesting\cite{fahmi1993theoretical,zhang2012computational}. However, there hasn't been much focus on reduced TiO$_2$  compounds like  Ti$_2$O$_3$. Previously,\cite{li2016investigation}, Yang Yang et al  observed a cubic perovskite  Ti$_2$O$_3$ unit cell made at room temperature at the Ti/SrTiO$_3$ interface, although they did not investigate the material's properties. We thoroughly investigate the structural, electrical, elastic, and optical properties of  Ti$_2$O$_3$ and SrTiO$_3$ materials for photovoltaic applications as a result of this.  Ti$_2$O$_3$ is Corundum structured and crystallizes in the trigonal R3c space group with Ti$^{3+}$ atoms bonded to six equivalent O$^{-2}$ atoms to form a mixture of edge, corner, and face-sharing TiO$_6$  octahedra as reported elsewhere \cite{guo2012electronic}. Similarly, SrTiO$_3$ is (Cubic) Perovskite structured and crystallizes in the tetragonal I4/mcm space group with Sr$^{2+}$ bonded to twelve O$^{-2}$ atoms to form SrO$_{12}$ cuboctahedra that share corners with twelve equivalent SrO$_{12}$ cuboctahedra. In this study, the stability and properties of these two compounds will be explored their suitability for use in solar harvesting applications.
Most recently\cite{allan2022impact}, the properties of titanium oxynitride compounds Ti$_n$N$_2$O$_{2n-1}$ with n=3, obtained  from pristine Ti$_2$O$_3$, were investigated. Ti$_2$O$_3$ displayed metallic properties, yet it had been reported to have semi-conducting properties experimentally\cite{aoki2019insulating}. The failure of first-principles DFT to properly describe the electronic band gaps of strongly correlated materials, in which the electron-electron interaction has a significant impact, is a well-known problem\cite{samat2016hubbard}. Many approaches to resolving this discrepancy have been proposed. The inclusion of the Hubbard model (DFT+U) in electronic structure calculations is one of them \cite{samat2016hubbard}. The DFT+U method has been used for improved prediction of the gaps as well as to exhaustively determine the elastic and mechanical properties of the compounds. The study of the elastic properties such as elastic constants provides fundamental information on how these materials behave under external strain\cite{burke2007abc}. Besides this, the effect of spin orbit coupling (SOC) has also been analyzed. To the best of our knowledge, there are a few experimental reports in the literature on the properties of Ti$_2$O$_3$ and SrTiO$_3$. Hence, this work aimed at investigating the properties of the two titanites containing Ti$^{3+}$ ions with a specific focus on their suitability for use in photovoltaic applications. Specifically, we explore the structural, electronic, elastic, mechanical, and optical properties of Ti$_2$O$_3$ and SrTiO$_3$ compounds using the DFT method with SOC and Hurbbard U effects. This study opens new doors towards some new titanite-based technologies.
This paper is organized as follows:  Sec.\ref{sec : II} accounts for the technicalities that may be needed for future reproducibility in our computations. We offer detailed analysis of results and discussions in Sec.\ref{sec:III}. The structural properties of both Ti$_2$O$_3$ and SrTiO$_3$ with and without SOC effects are discussed in Sec.\ref{sec : IIIA} while the eEffect of SOC and Hubbard U on electronic Properties in Sec.\ref{sec : IIIB}. The elastic constants and mechanical properties of the compounds under study are outlined in  Sec.\ref{sec : IIIC}.  In Sec.\ref{sec : IIID}, we discuss the optical properties of Ti$_2$O$_3$ and SrTiO$_3$.    Sec.\ref{sec :IV} outlines the conclusions of this work.

\section{Computational details}\label{sec : II}

The structural, electronic, elastic, mechanical, and optical properties of  Ti$_2$O$_3$ and SrTiO$_3$ compounds were computed within the DFT\cite{burke2007abc,Kohn-65} as implemented in the quantum espresso (QE) code \cite{Giannozzi_2009}. The Ti$_2$O$_3$ and SrTiO$_3$ crystal structure input files were downloaded from the materials project database \cite{jain2013commentary} and the materials cloud \cite{talirz2020materials} input generator implemented in QE was used to generate the PWscf input files for DFT calculations. The electron-ion interaction was denoted by using the projector augmented-wave function (PAW) method\cite{PAW-PPS}. The Generalized Gradient Approximation (GGA) with the Perdew-Burke-Ernzerhof (PBE) \cite{PBE-1996} was chosen to define the exchange-correlation effect of the electrons. The optimized cutoff energy of 50Ry and 6 × 6 × 4 Monkhorst-Pack grid for Brillouin zone integration were used for the calculations. Geometry optimization was performed by computing the total energy per unit cell at several lattice constant values to obtain the ground state structural properties. Based on the optimized lattice constants, the elastic, electronic, and optical properties were calculated. The effects of spin orbit coupling (SOC) was included in the structural and electronic properties, while the well-known DFT underestimation problem was countered but the inclusion of Hubbard U correction on the electronic structure calculations. The choice of the effective Hubbard U values were peged on earlier works done on same materials while assesing the impact of the inclusion of U on the lattice parameter(a$_0$) of the compounds.

\section{Results and Discussions}\label{sec:III}

\subsection{Structural Properties}\label{sec : IIIA}

\begin{figure}[t]
    \centering
    \includegraphics[width=\textwidth]{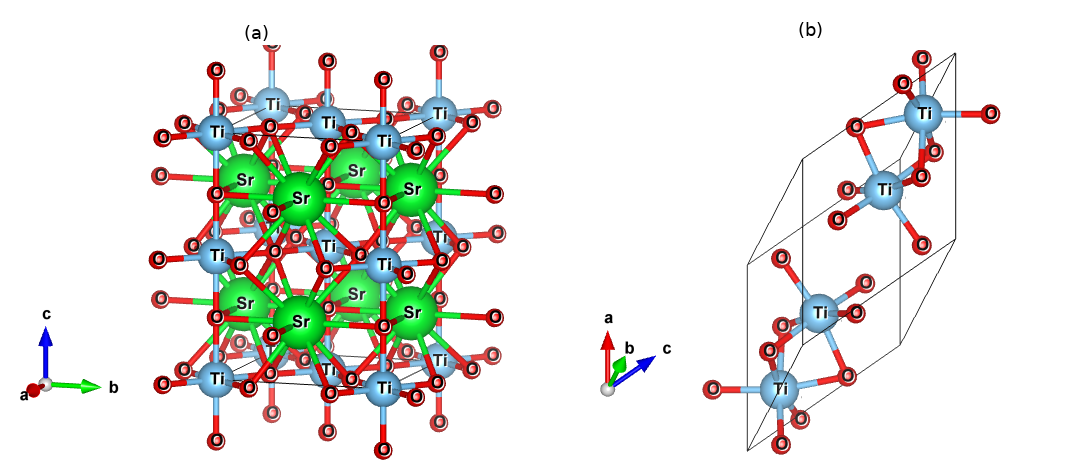}
    \caption{Optimized Crystal structure for (a)  SrTiO$_3$ and (b) Ti$_2$O$_3$ compounds}
    \label{fig:fig1}
\end{figure}

Ti$_2$O$_3$ and SrTiO$_3$ compounds adopt trigonal and tetragonal crystal system as reported elsewhere \cite{talirz2020materials}. In order to achieve the equilibrium structure, one has to calculate the lattice parameter that minimizes the DFT total energy, the optimized crystal structures are shown in Figure \ref{fig:fig1}. Table \ref{tab:table1} shows bond strength for  Ti$_2$O$_3$ and SrTiO$_3$.

\begin{table}[t]
\begin{tabular}{lll} 
 Material                & Bond  & Bond length \\ \hline\hline
\multirow{2}{*}{Ti$_2$O$_3$}  & Ti-O  & 2.04        \\ 
                        & Ti-Ti & 2.37        \\ \hline
\multirow{3}{*}{SrTiO$_3$} & Sr-Ti & 3.41        \\ 
                        & Sr-O  & 2.78        \\ 
                        & Sr-Sr & 3.95        \\ \hline\hline
\end{tabular}
\caption{Bond lengths for Ti$_2$O$_3$  and SrTiO$_3$  compounds}\label{tab:table1}

\end{table}

\begin{figure}[]
    \centering
    \includegraphics[width=0.9\textwidth]{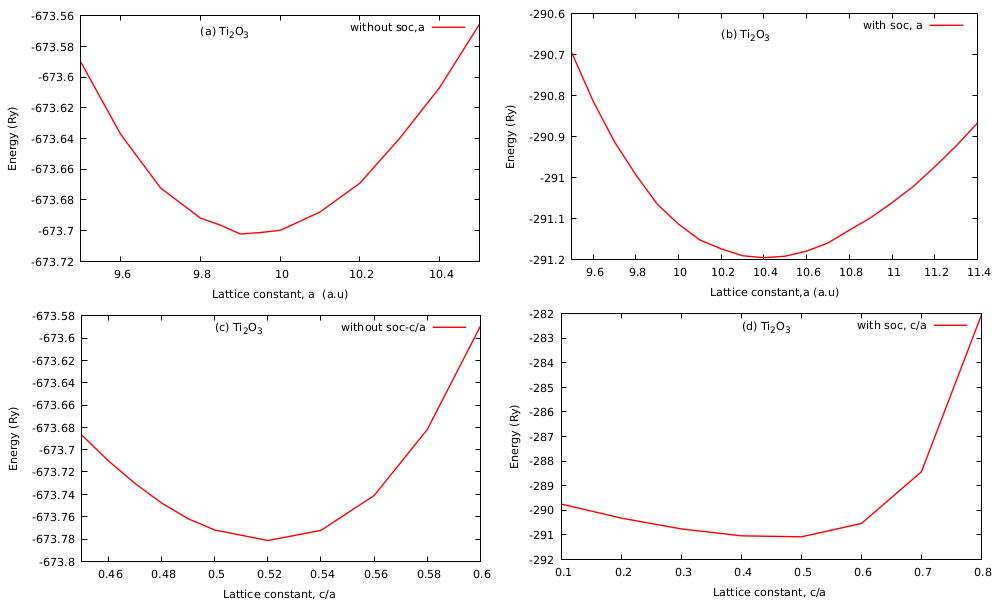}
    \caption{Total energy versus lattice constants computed with and without SOC effects for Ti$_2$O$_3$ compound.}
    \label{fig:fig2}
\end{figure}

\begin{figure}[]
    \centering
    \includegraphics[width=0.9\textwidth]{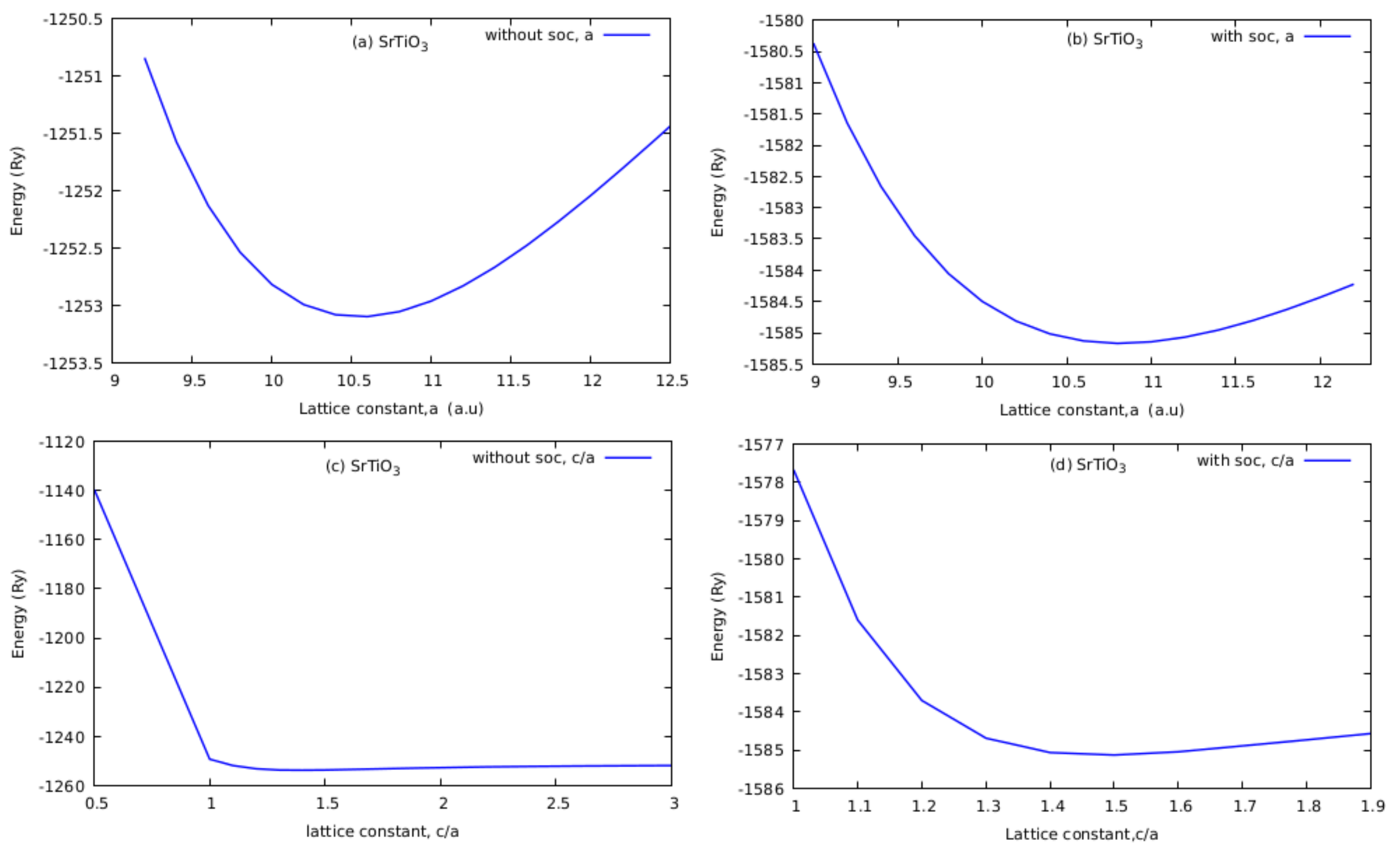}
    \caption{Total energy versus lattice constants computed with and without SOC effects for SrTiO$_3$  compound.}
    \label{fig:fig3}
\end{figure}

From Table \ref{tab:table1}, the Ti-O bond is stronger than the Ti-Ti bond in Ti$_2$O$_3$, while the Ti-O bond is stronger than both Ti-Sr and Sr-O bonds in SrTiO$_3$ . The Sr-Sr bond is the weakest with long bond lengths; this difference is attributed to the fact that the volume per atom tends to increase with the increase in atomic radius and therefore affects crystal lattice basis\cite{mooser1956chemical}. The total energy at various lattice constant values is computed with SOC and without SOC effects are presented in figure \ref{fig:fig2} and figure \ref{fig:fig3} for Ti$_2$O$_3$  and SrTiO$_3$  compounds, respectively. Table \ref{tab:table2} gives a summary of the calculated lattice parameters and other structural properties.

\begin{table*}[]
\begin{tabular}{lllll}
\hline\hline
Structural Properties                                                        & \multicolumn{2}{l}{Ti$_2$O$_3$} & \multicolumn{2}{l}{SrTiO$_3$ } \\ 
                           & Without soc & With  soc & Without soc & With soc \\ \hline\hline
Lattice Parameters a$_0$(a.u) & 9.933       & 10.404    & 10.535      & 10.849   \\ \hline
Lattice parameter c/a      & 0.520       & 0.571     & 1.353       & 1.403    \\ \hline
Bulk modulus B$_o$ (GPa)      & 166.4       & 137.0     & 217.1       & 218.2    \\ \hline
\begin{tabular}[c]{@{}l@{}}Ground state energy\\    \\ E$_o$ (ry)\end{tabular}  & -673.7     & -291.1       & -1253.07     & -1585.16    \\ \hline
\begin{tabular}[c]{@{}l@{}}Equilibrium volume\\    \\ V$_o$ (a.u)3\end{tabular} & 980.14     & 1126.09     & 1167.17      & 1277.00     \\ \hline\hline
\end{tabular}
\caption{Effect of SOC on Structural properties of Ti$_2$O$_3$ and SrTiO$_3$ compounds} \label{tab:table2}
\end{table*}

From figure \ref{fig:fig2} and figure \ref{fig:fig3}, SOC did not have significant effects on the lattice parameters. However, when the data was fitted to Murnaghan’s equation of state\cite{morgan2010intrinsic}, minimum changes were noted on the equilibrium volume and ground state energies of the two compounds as indicated in table \ref{tab:table2}

\subsection {Effect of SOC and Hubbard U on Electronic Properties}\label{sec : IIIB}

\begin{figure}[b]
    \centering
    \includegraphics[width=0.9\textwidth]{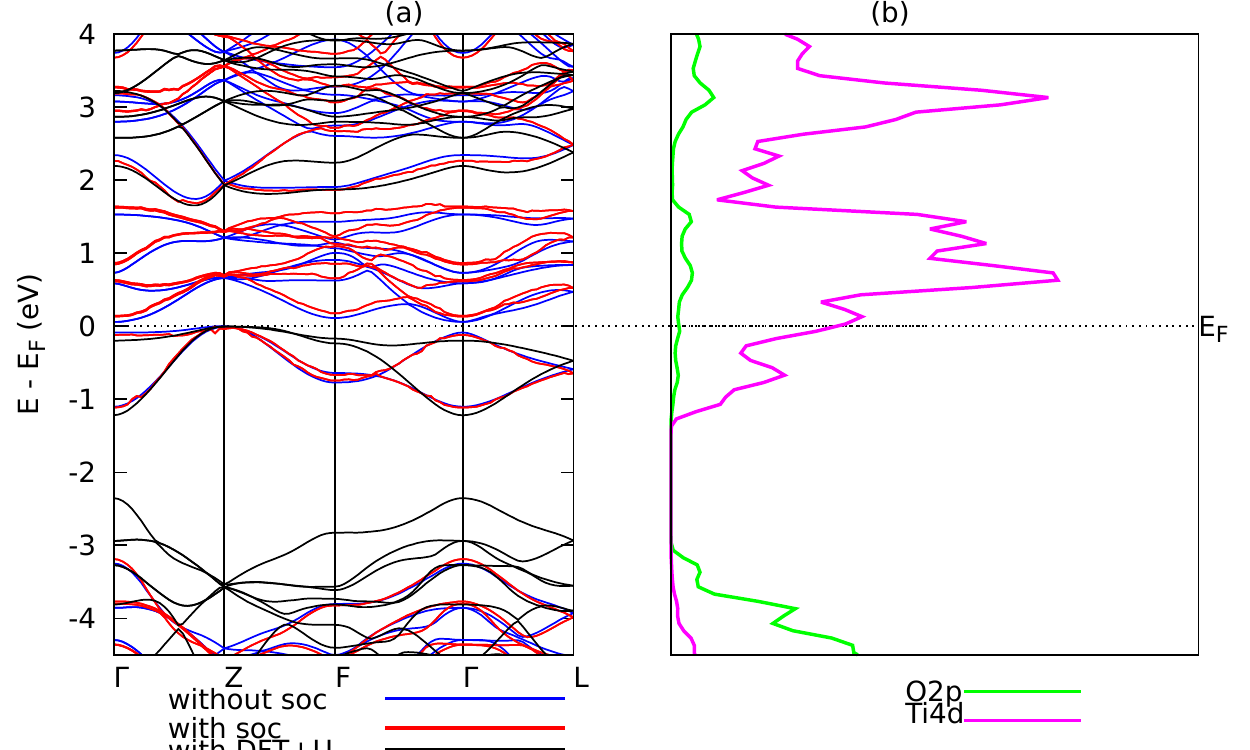}
    \caption{The calculated (a) band structures of Ti$_2$O$_3$ compounds without SOC effects(blue), with SOC(red) and with DFT+U (U=5 eV) (black), (b)The PDOS for Ti$_2$O$_3$  compound without SOC and without DFT+U effects.}
    \label{fig:fig4}
\end{figure}

\begin{figure}[b]
    \centering
    \includegraphics[width=0.9\textwidth]{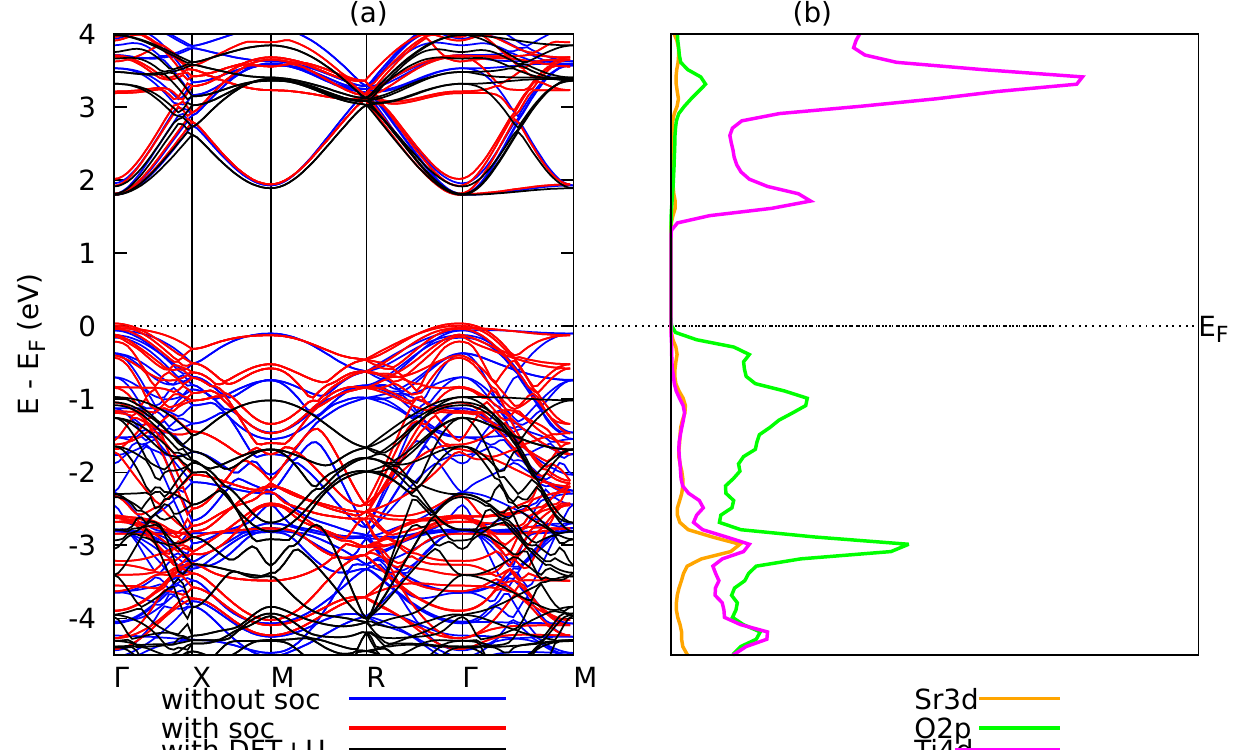}
    \caption{The calculated (a) band structures of SrTiO$_3$ compounds without SOC effects (blue), with SOC(red) and with DFT+U (U=5 eV) (black), (b)The PDOS for SrTiO$_3$  compound without SOC and without DFT+U effects.}
    \label{fig:fig5}
\end{figure}

The electronic band structures and projected density of states (PDOSs) of the Ti$_2$O$_3$  and SrTiO$_3$ compounds were computed by using the optimized crystal structures and presented on figure \ref{fig:fig4} and figure \ref{fig:fig5}. The band structures were calculated in three stages; i) without SOC effects, ii) with SOC effect, and iii) with Hubbard U corrections. The choice of U values were pegged on the earlier works on the same materials by ref\cite{allan2022impact} and ref\cite{adewale2019first} for Ti$_2$O$_3$  and SrTiO$_3$ , respectively. For Ti$_2$O$_3$  and SrTiO$_3$ , Hubbard U=5 eV and U=7 eV, respectively, proved to have no significant effects on the lattice constants and were therefore used for the band structure calculations.

The Ti$_2$O$_3$  and SrTiO$_3$  compounds have narrow bandgaps of Ti$_2$O$_3$  has been found to have electronic bandgaps of 0.059 eV and 1.612 eV, respectively, without SOC, 0.131 eV and 1.761 eV, respectively, with SOC, and 1.665 eV and 2.769 eV, respectively, with Hubbard U as seen in figures \ref{fig:fig4} and \ref{fig:fig5}(a). The summary of the calculated bandgaps is tabulated in table \ref{tab:table3}. 

\begin{table*}[]
\begin{tabular}{llllll}
\hline\hline
Materials & Calculated Band Gaps &&& Experimental gaps &                    \\ \hline
          & Without SOC          & With SOC          & With Hubbard U &              \\ \hline
 Ti$_2$O$_3$   & 0.0598               & 0.1311            & 1.667 (U=5eV)  & 1.67\cite{li2016investigation} \\ 
SrTiO$_3$   & 1.612                & 1.7612            & 2.769(U=7eV)   & 2.85\cite{van2001bulk} \\ \hline\hline
\end{tabular}
\caption{The calculated and experimental bandgaps}\label{tab:table3}
\end{table*}

From the band structure calculations, SOC did not have significant effect on the bandgaps, however the DFT+ U improved the prediction of the band gaps with respect to the experimental values. The maxima of the valence bands and the minima of the conduction bands occur at different symmetry points (Z-$\Gamma$) for  Ti$_2$O$_3$  and SrTiO$_3$  and ($\Gamma$–M) for SrTiO$_3$  in the Brillouin zone, implying that the two compounds are indirect bandgap semiconductors. The projected density of states, which describes the available states for electrons to occupy when projected on atomic orbitals were calculated. However, since the effects of SOC and without Hubbard U did not have significant impact on the types of dominant states in the PDOS, only the results of the PDOS without the effects of SOC and Hubbard U are presented in figures \ref{fig:fig4}(b) and \ref{fig:fig5}(b). The orbital contributions to the formation of valence bands and conduction bands are described by the PDOS in the energy region -4.5 eV to 4 eV. As illustrated in figure \ref{fig:fig4}(b), the entire band structure is dominated with O-2p and Ti-4d orbitals, an indication that  Ti$_2$O$_3$ exhibits semi metallic properties without Hubbard U effects, this is attributed to the well-known poor description of the electronic structure by the DFT method. The bands in the region -4.5 to -2.5 eV are majorly formed by O-2p while Ti-4d states forms the bands in the energy region of -1.0 eV to 4 eV in the Ti$_2$O$_3$ band structure shown in figure  \ref{fig:fig5}. In the case of SrTiO$_3$  compound (Figure \ref{fig:fig5}(b)), the valence band is majorly formed by the O-2p states in the energy range of -4.5eV to Fermi level, while the conduction band is majorly formed by the Ti-4d orbitals.

\subsection{Mechanical and Elastic properties} \label{sec : IIIC}

The Ti$_2$O$_3$  and SrTiO$_3$ compounds adopt trigonal and tetragonal  crystal structure featuring six independent  elastic constants. The independent elastic constants for trigonal Ti$_2$O$_3$  are C$_{11}$  C$_{12}$  C$_{13}$ C$_{14}$  C$_{33}$ C$_{44}$  \cite{chung1967voigt}.
The necessary and sufficient conditions for trigonal elastic  Ti${_2}$O${_3}$  stability is given by equation(1);

\begin{align}
    \begin{split} \label{eqn1} 
      C_{11}> \left | C_{12} \right |;C_{44}> 0;  \\C_{13}^{2}< \frac{1}{2}C_{33}\left (C_{11} +C_{12}\right);\\C_{14}^{2}< \frac{1}{2}C_{44}\left ( C_{11} -C_{12}\right )= C_{44}C_{66};\\C_{66} = \frac{1}{2}(C_{11}- C_{12})
    \end{split}
\end{align}

For Tetragonal  SrTiO$_3$ , the six independent elastic constants are C$_{11}$  C$_{12}$  C$_{13}$  C$_{14}$  C$_{33}$   C$_{44}$  and C$_{66}$ \cite{chung1967voigt}. The necessary and sufficient conditions for elastic stability of the tetragonal crystal system are ;

\begin{align}
    \begin{split} \label{eqn2} 
      C_{11}> \left | C_{12} \right |;C_{44}> 0;\\ 2C_{13}^{2}< C_{33}\left ( C_{11} + C_{12}\right ); C_{66}> 0  
    \end{split}
\end{align}

Table \ref{tab:table4} shows the calculated elastic constants for both the compounds, from the results, both the compounds meet the stability criteria, at DFT and DFT+U levels, and are therefore mechanically stable. The bulk modulus B, Young’s modulus E, shear modulus G , Pugh’s ratio B/G, and Poisson’s ratio n are presented in table \ref{tab:table5} .

\begin{table*}[]
\caption{Computed elastic constants C$_{ij}$ (GPa) of Ti$_2$O$_3$  and SrTiO$_3$  compounds} \label{tab:table4}
\begin{tabular}{lllllllll}
\hline\hline
Compound                &       & C$_{11}$   & C$_{12}$   & C$_{13}$   &C$_{14}$  & C$_{33}$   &C$_{44}$   &C$_{66}$  \\ \hline
\multirow{2}{*}{ Ti$_2$O$_3$}  & DFT   & 350.3 & 138.9 & 175.7 & 83.04 & 276.0 & 114.7 & 105.6 \\  
                        & DFT+U & 285.3 & 102.1 & 113.4 & 3.073 & 228.4 & 121.5 & 91.6  \\ \hline
\multirow{2}{*}{SrTiO$_3$} & DFT   & 338.3 & 110.6 & 105.7 & -     & 351.6 & 116.5 & 122.3 \\ 
                        & DFT+U & 372.5 & 116.7 & 115.2 & -     & 383.3 & 130.3 & 133.0 \\ \hline\hline
\end{tabular}

\end{table*}

\begin{table}[]

  \begin{tabular}{lllllll}
  \hline\hline
Compound &       & B     & E     & G     & B/G  & n    \\ \hline
Ti$_2$O$_3$     & DFT   & 217.0 & 250.2 & 95.7  & 2.26 & 0.31 \\ 
         & DFT+U & 161.3 & 239.4 & 95.5  & 1.69 & 0.25 \\ \hline
SrTiO$_3$   & DFT   & 185.8 & 292.4 & 118.1 & 1.57 & 0.24 \\ 
         & DFT+U & 202.5 & 322.9 & 130.8 & 1.54 & 0.23 \\ \hline\hline
\end{tabular}
\caption {Mechanical properties of  Ti$_2$O$_3$  and SrTiO$_3$  compounds} \label{tab:table5}
\end{table}

Bulk modulus measures the resistance against volume change resulting from applied external pressure \cite{allan2022first}. Large B value predicts hard materials; thus, from the computed values of the bulk modulus, both Ti$_2$O$_3$  and SrTiO$_3$  are not hard materials. It is clear that for Ti$_2$O$_3$, DFT+U gives a more accurate prediction of the mechanical properties with reference to the values obtained (in table \ref{tab:table2}) when lattice constants were fitted to Murnaghan’s equation of state\cite{murnaghan1944compressibility}. Additionally, the bond lengths of the crystal structures are correlated to the size of the B. The shorter the bond lengths, the larger the B value \cite{manyali2012ab}. From the structural properties, the obtained bond lengths in Ti$_2$O$_3$ are shorter than those in the  SrTiO$_3$ crystal structure thus the higher value of B in Ti$_2$O$_3$ compound by DFT calculations. The ductile (ionic) and brittle (covalent) nature of materials is determined by Pugh’s ratio B/G and Poisson’s ratio, n \cite{allan2022first}. The restriction for brittleness is B/ G $<$ 1.75; otherwise, the material is said to be ductile. In addition, the Cauchy pressure (C$_{12}$ -C$_{44}$ ) also confirms the ductility and brittleness of materials. A positive value of the Cauchy pressure indicates ductility and brittleness otherwise. The critical value for the Poisson’s ratio is 0.26, with materials having n as 0.26 exhibiting a ductile nature and brittleness otherwise. From the calculated values, all the three methods are consistent in classifying the materials as brittle. Poisson’s ratio n  0.1, 0.25, and 0.33 for pure covalent, ionic, and metallic bonds, respectively, \cite{manyali2012ab}. Thus, we can conclude that Ti$_2$O$_3$  and SrTiO$_3$ compounds are ductile and strongly dominated by ionic character. The stiffness of a material is determined by applying Young’s modulus value \cite{manyali2012ab}. The higher the value of E, the stiffer the material, therefore, SrTiO$_3$  compound is stiffer than Ti$_2$O$_3$.

\subsection {Optical Properties}\label{sec : IIID}

Optical properties of solids, which dependent on their band gap, are very interesting to know how the solid interacts with light. To highlight optical properties such as: reflection transmission and absorption, we need to calculate the dielectric function $\varepsilon \left ( \omega  \right )$, which reflects the response of electrons in solid to electromagnetic irradiation. The dielectric function $\varepsilon \left ( \omega  \right )$ is given by :

\begin{equation}\label{eqn3}
  \varepsilon \left ( \omega  \right ) = \varepsilon_{1}\left ( \omega  \right ) +
i\varepsilon_{2}\left ( \omega  \right )
\end{equation}

\begin{figure*}[t]
     \centering
     \includegraphics[width=0.9\textwidth]{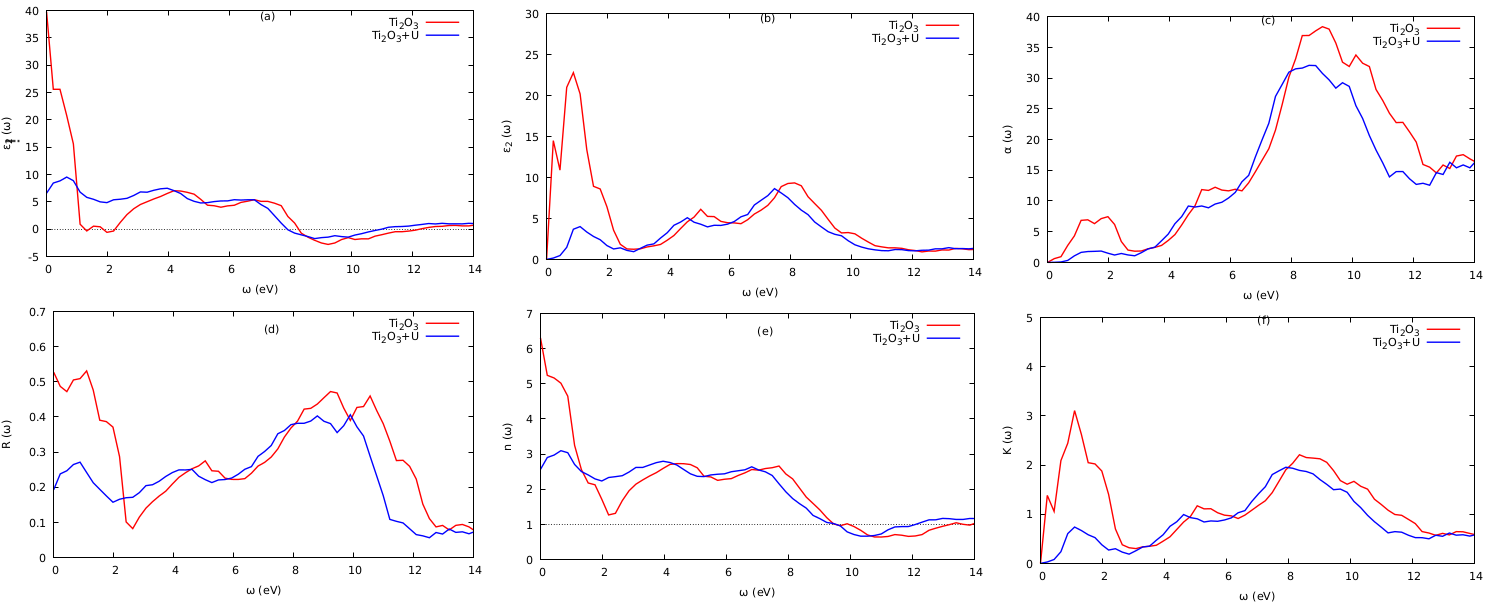}
     \caption{Optical Properties for  Ti$_2$O$_3$}
     \label{fig:Fig9}
 \end{figure*}

\begin{figure*}[t]
     \centering
     \includegraphics[width=0.9\textwidth]{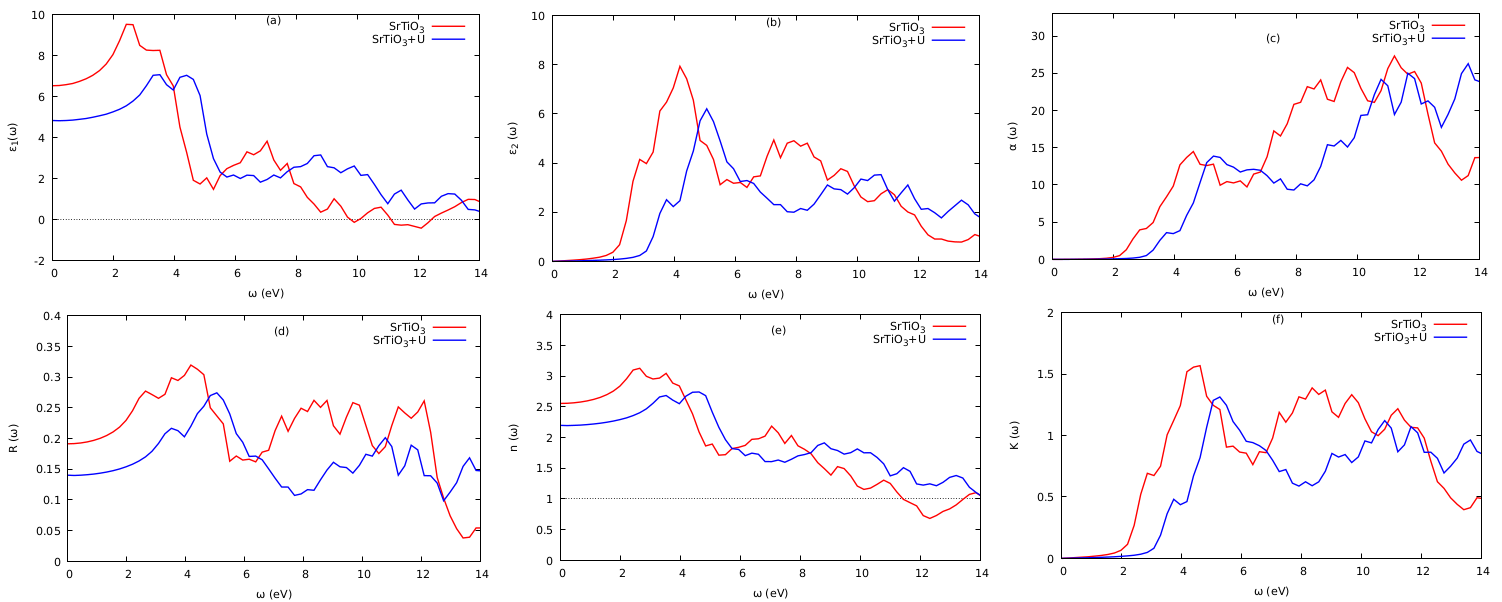}
     \caption{Optical Properties for  SrTiO$_3$}
     \label{fig:Fig10}
 \end{figure*}

\begin{figure*}[t]
     \centering
     \includegraphics[width=0.9\textwidth]{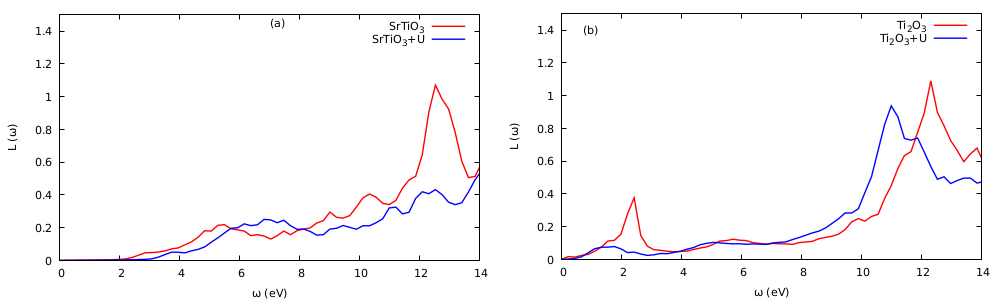}
     \caption{Energy loss $L\left ( \omega  \right )$  for  ATiO$_3$}
     \label{fig:Fig11}
 \end{figure*}

$\varepsilon_{1}\left ( \omega  \right )$ and $\varepsilon_{2}\left ( \omega  \right )$ are the real and the imaginary parts of the dielectric function. The real part is associated to the electronic polarizability of the material while the imaginary part is correlated to the electronic absorption of material\cite{aslam2021structural}.  The real part and the imaginary parts describe, respectively, the dispersion and the absorption of the electromagnetic radiation by the medium which it crosses. The spectrum is calculated by summing the electric dipole operator matrix elements between the occupied and unoccupied wave functions over the Brillouin zone while respecting the selection rules. This is mainly connected with the electronic structures and characterizes the linear response of the material to electromagnetic radiation. It therefore governs the propagation behavior of radiation in a medium. The imaginary part of the dielectric function represents the electron transition between the valence and the conduction bands. The other optical properties like as: the reflectivity $R\left ( \omega  \right )$ , the absorption $\alpha\left ( \omega  \right )$ and the refractivity $n\left ( \omega  \right )$, can be derived from $\varepsilon_{1}\left ( \omega  \right )$  and $\varepsilon_{2}\left ( \omega  \right )$. These are presented in Figure \ref{fig:Fig9},  \ref{fig:Fig10}  and  \ref{fig:Fig11} are computed by using the equations presented as follows \cite{samat2019structural,aslam2021structural,arbab2019optical}:

\begin{equation} \label{eqn4}
  \alpha\left( \omega \right) = \sqrt{2}\omega\left( \sqrt{\varepsilon_{1}^{2}\left( \omega \right) + \varepsilon_{2}^{2}\left( \omega \right)} - \varepsilon_{1}\left( \omega \right) \right)^{1/2}                                                                                                         
\end{equation} 

\begin{equation}\label{eqn5}
{n\left( \omega \right) = \sqrt{2}\omega\left( \frac{\sqrt{\varepsilon_{1}^{2}\left( \omega \right) + \varepsilon_{2}^{2}\left( \omega \right)} - \varepsilon_{1}\left( \omega \right)}{2} \right)^{1/2}}              
\end{equation}

\begin{equation} \label{eqn6}
 K\left( \omega \right) = \sqrt{2}\omega\left( \frac{\sqrt{\varepsilon_{1}^{2}\left( \omega \right) + \varepsilon_{2}^{2}\left( \omega \right)} - \varepsilon_{1}\left( \omega \right)}{2} \right)^{1/2}                                                                                                          
\end{equation}

\begin{equation} \label{eqn5}
 L\left( \omega \right) = \frac{\varepsilon_{2}(\omega)}{\varepsilon_{1}^{2}\left( \omega \right) + \varepsilon_{2}^{2}(\omega)}   
\end{equation}

\begin{equation} \label{eqn7}
R\left( \omega \right) = \frac{{(n - 1)}^{2} + K^{2}}{{(n + 1)}^{2} + K^{2}}
\end{equation}

The imaginary part ($\varepsilon_{2}\left ( \omega  \right )$) of the dielectric wave function describes the photon absorption in crystalline materials\cite{guo2012electronic}. The peaks in ($\varepsilon_{2}\left ( \omega  \right )$)curves result from the electronic transitions from the valence to the conduction bands. The absorption onsets in $\alpha\left ( \omega  \right)$ curves refer to the materials bandgaps which lie within the visible region, $<$3.1 eV for for Ti$_2$O$_3$  and SrTiO$_3$ . compounds, and are consistent with the bandgaps from the band structure, an implication of strong inter-band transitions. This makes for Ti$_2$O$_3$  and SrTiO$_3$ compounds promising candidates for photovoltaic applications. Additionally, narrow bandgaps facilitate faster electron transitions as opposed to wide bandgaps \cite{samat2019structural}. The key feature of the($\varepsilon_{1}\left ( \omega  \right )$) curve is($\varepsilon_{1}\left ( \omega  \right )$) Energy=0), also referred to as the static value\cite{arbab2019optical}. This static value is correlated to the material’s refractive index as n ($\omega  $) . Starting from Energy 0, the ($\varepsilon_{1}\left ( \omega  \right )$) plot attained major peaks at low energy regions, $<$1.5eV and $<$2.7 eV for Ti$_2$O$_3$  and SrTiO$_3$, respectively. The photon transmission persisted until the Re($\omega$) values became negative at energy regions of 8eV - 12eV. At this energy region, the incident photon radiations are assumed to be fully attenuated\cite{murtaza2011first}  and the compounds assert a metallic behaviour\cite{murtaza2011investigation}. From n ($\omega  $) curves, the calculated refractive indices obtained at zero energies are 2.3 (DFT$+$U) and 2.5 (without U) for  SrTiO$_3$  and 2.5 (with Hubbard U) for  SrTiO$_3$ . The major refractive index peaks for SrTiO$_3$ and Ti$_2$O$_3$ (with Hubbard U) reside within the visible region. Additionally, the Ti$_2$O$_3$  and SrTiO$_3$  were found to have high optical absorption in the energy range of 8eV-10eV and 10eV-12eV, respectively. The optical absorption coefficients of Ti$_2$O$_3$  and SrTiO$_3$ compounds cover the UV-Vis regions. The absorption coefficients calculated covered a wide range of the electromagnetic spectrum in the energy regions 2.0–13.5 eV; this demonstrates that these compounds can be utilized for photovoltaic applications. The materials’ surface behaviour and energy loss by fast electrons entering a medium are determined by reflectivity and energy loss function, respectively \cite{arbab2019optical}. The main peaks of the reflectivity curves are observed in the regions 6 eV -11 eV and 2 eV-6 eV for Ti$_2$O$_3$  and SrTiO$_3$, respectively, the reflectivity decreased beyond this region. There was no significant absorption in the visible regions, as depicted in the loss spectrum for both compounds in figures  \ref{fig:Fig11}(a) and  \ref{fig:Fig11}(b). The major absorption peak occurred at higher energy regions 9.97–12.71 eV. The major absorption peak occurred at higher energy regions $>$12 eV and  $>$10 eV for SrTiO$_3$ and Ti$_2$O$_3$, respectively. The optical properties  results obtained in this work are in agreement with the results obtained previously on the related materials\cite{samat2019structural,aslam2021structural}, .

\section{Conclusion} \label{sec :IV}

We have studied the structural, electronic, elastic, mechanical, and optical properties of Ti$_2$O$_3$  and SrTiO$_3$ .compounds using the DFT (with and without SOC effects), and DFT+U methods as implemented in the QE package. Interestingly, SOC did not have significant effects on the structural and electronic properties of the compounds. Equilibrium lattice constants of 9.93 a.u without SOC and 10.40 a.u with SOC were obtained for Ti$_2$O$_3$ compound, whereas for SrTiO$_3$  the calculated equilibrium lattice constants were 10.53 a.u and 10.84 a.u without SOC and with SOC, respectively . Ti$_2$O$_3$ was found to have electronic bandgaps of 0.059 eV without SOC and 0.131 eV with SOC while  SrTiO$_3$ electronic bandgaps were 1.612 eV, and 1.761 eV, respectively, without SOC and with SOC. When the Hubbard U effects were in cooperated in the electronic structure calculations, the bandgaps for Ti$_2$O$_3$  and SrTiO$_3$, were predicted as 1.665 eV and 2.769 eV, respectively. These were in fairly good agreement with reported experimental results. The formation of the conduction band was primarily by Ti-4d, while the valence bands were dominated by O-2p orbitals in both compounds. Both Ti$_2$O$_3$  and SrTiO$_3$  were found to be mechanically stable at zero pressure, ductile, and ionic, thus their potentiality for resilient materials application, these properties were confirmed by DFT+U studies. Additionally, the optical properties of Ti$_2$O$_3$  and SrTiO$_3$ were improved by DFT+U studies. The calculated bandgaps, high refractive indices, high absorption coefficients, and wide energy coverage of the absorption coefficients spectra was mainly in the UV-Vis regions of the electromagnetic spectrum.  This work suggests that Ti$_2$O$_3$  and SrTiO$_3$ compounds are suitable for photovoltaic applications.

\section{Acknowledgments}

This work was supported by the Partnership for Skills in Applied Sciences, Engineering and Technology (PASET)— Regional Scholarship Innovation Fund (RSIF). The authors acknowledge the Centre for High-Performance Computing, CHPC, Cape for HPC resources.

\section{Data availability}

The source files and  data will be assessed upon request from the authors.

\section{conflicts of interest}

The authors declare that they have no conflicts of interest.

\bibliography{lynn}

\end{document}